# Regroupement sémantique de définitions en espagnol

Gerardo Sierra[1], Juan-Manuel Torres-Moreno[2,3], Alejandro Molina[2]

[1]Universidad Nacional Autónoma de México, Ciudad Universitaria, México D.F.
gsierram@iingen.unam.mx     http://www.iling.unam.mx

[2]Laboratoire Informatique d'Avignon, BP1228, 84911 Avignon Cédex 09, France
juan-manuel.torres@univ-avignon.fr     alejandro.molina@etd.univ-avignon.fr
http://lia.univ-avignon.fr/

[3]École Polytechnique de Montréal, CP 6079 Succ. Centre-ville, Montréal (Québec) Canada

**Résumé.** Cet article s'intéresse à la description et l'évaluation d'une nouvelle méthode d'apprentissage non supervisé pour réunir des définitions en espagnol selon leur signification. Nous utilisons comme mesure de regroupement l'énergie textuelle et nous étudions une adaptation de la précision et le rappel afin d'évaluer notre méthode.

## 1   Introduction

De nos jours, l'utilisation de l'Internet pour la recherche de définitions est de plus en plus importante. Wikipédia et Medline sont devenu les sites les plus consultés de la Web[1]. Or, il existe un énorme nombre de définitions qui sont parfois inaccessibles aux utilisateurs. Celles-ci peuvent se trouver dans des sites non encyclopédiques ou dans de documents divers. Dans cette perspective nous avons développé le moteur de recherche *Describe*[2], qui permet de trouver des définitions en espagnol (Sierra et al., 2009). Une caractéristique de ce moteur est qu'il regroupe les résultats des recherches (définitions liées à un terme). Cet article présente la méthodologie de regroupement et l'évaluation des résultats. Ceux-ci sont encourageants du point de vue qualitatif. Par contre, l'évaluation quantitative pose des contraintes car il est compliqué d'évaluer la sémantique.

Cet article est organisé comme suit : dans la section 2 nous introduisons les contextes définitoires (CD), dans la section 3 nous présentons des stratégies de regroupement des définitions. Le corpus utilisé dans nos expériences est présenté en section 4. Des évaluations avec des analyses quantitative et qualitative sont présentées au chapitre 5 avant de conclure et de donner quelques perspectives.

## 2   Contextes définitoires

Un contexte définitoire (CD) est un fragment textuel qui définit un terme (Alarcón et al., 2009). Par exemple: *Le navire, peut être originellement considéré comme un flotteur qui*

---
[1] http://www.alexa.com/topsites
[2] http://www.describe.com.mx

*essaie de rester dans une position verticale face à des perturbations extérieures*. Le fragment précédent est un CD car il possède le terme (T) : *navire* et sa définition respective (D) : *un flotteur qui essaie …* Dans cet exemple, le terme et sa définition se rattachent grâce à la structure syntaxique *peut être+considéré+comme*. Ce type de structure est nommée patron définitoire (PD). Mis à part le terme et la définition, un CD possède d'autres éléments ayant déjà été définis dans d'autres travaux (Sierra et al., 2004). Dans cet article il suffit de connaître les deux éléments auparavant décrits.

Les PDs peuvent être des structures syntaxiques bien spécifiques, comme la séquence : *peut+se+définir+comme*[3] (Rodríguez, 2004). Chaque PD peut s'associer à un type de définition établie selon la relation entre T et D. L'extraction de CDs moyennant des PDs en français a été étudiée par Malaisé (2005) et Rebeyrolle et Tanguy (2000). En espagnol, il y a quatre types de définitions basées sur le modèle aristotélique : l'analytique, l'extensionnel, le fonctionnel et le synonymique. Il existe des études approfondies des trois premiers (Aguilar et Sierra, 2009), pour cette raison nous laissons de côté le modèle synonymique.

Les définitions analytiques présentent un genus qui exprime la catégorie la plus générale à laquelle le terme appartient et une différentia qui permet de le distinguer d'autres éléments de la même catégorie. Quelques patrons verbaux liés à ce type de définition sont : *être+un, définir+comme, comprendre+comme*. Les définitions extensionnelles énumèrent les parties qui conforment le terme défini. Quelques verbes liés aux définitions extensionnelles sont : *contenir, comprendre,* et *inclure*. Les définitions fonctionnelles ne présentent pas un genus, mais une différentia qui exprime l'usage particulier du terme est introduite. Quelques patrons liés à ce type de définitions sont : *fonctionner, permettre* et *sert+à*.

## 3 Regroupement de définitions

### 3.1 Regroupement hiérarchique

On introduira d'abord quelques définitions nécessaires à la compréhension du reste de la section. Un document est une chaîne fini de longueur arbitraire de symboles graphiques nommés entités lexicales (EL). Une EL est soit un symbole soit l'union de plusieurs symboles. Par exemple, le mot *pomme* ou un groupe de mots comme *République Française*. De même, une EL peut être un symbole inintelligible comme *Viv* ou *B4*. Une collection est un ensemble de documents et un dictionnaire est la liste de ELs uniques apparaissant dans une collection.

La représentation des définitions est basée sur le modèle vectoriel de Salton (1971). Les définitions sont des vecteurs avec le même nombre de dimensions que d'ELs dans le dictionnaire. Néanmoins, étant donné que les définitions sont généralement courtes, nous avons utilisé l'approche présence/absence de termes : 1 si l'EL apparaît et 0 autrement. La collection des définitions peut être donc représentée par une matrice binaire document-EL. Evidemment, cette représentation empêche l'utilisation de mesures de similarité basées sur des pondérations réelles (cosinus par exemple).

Le regroupement hiérarchique offre un grand avantage : le nombre de classes ne doit pas être spécifié. Dans la version la plus simple (HAC) l'entrée est un ensemble d'objets et la sortie un dendrogramme, c'est-à-dire un arbre hiérarchique qui regroupe tous les objets.

---

[3] Traduction en français de *se+puede+definir+como*.

Dans chaque itération, une fonction calcule la distance entre chaque paire de groupes afin de déterminer les deux groupes à fusionner. Le critère pour calculer la distance entre chaque groupe est généralement une variante des méthodes *linkage*. Dans la méthode du *complete linkage* (Sorensen, 1948), la distance entre les groupes est représentée par la plus grande distance entre un objet du premier groupe et un objet du deuxième. Cette méthode a l'avantage de créer de petits groupes, cohésifs et bien délimités. C'est la raison principale pour laquelle nous l'avons utilisée. La distance entre le groupe $D_i$ et le groupe $D_j$, où $d_i \in D_i$ et $d_j \in D_j$, est définie par :

$$Dist(D_i, D_j) = \max_{d_i \in D_i, d_j \in D_j} dist(d_i, d_j) \tag{1}$$

où *dist* est une fonction dont l'ensemble de départ sont les objets, à l'opposé de *Dist* dont l'ensemble de départ sont les groupes.

### 3.2 La valeur de seuil par distance

Pour obtenir un ensemble de groupes, nous utilisons un seuil, afin de stopper les fusions du dendrogramme. De cette façon, on obtient les groupes dont la distance se trouve en dessous d'un seuil prédéterminée $\alpha$. À chaque itération on détermine si les groupes $D_i$ et $D_j$ sont suffisamment proches, c'est-à-dire, si $Dist(D_i,D_j) \leq \alpha$. Dans le cas contraire, l'algorithme s'arrête et on garde les regroupements effectués.

Nous n'avons pas encore spécifié la façon de calculer la distance $dist(d_i,d_j)$. Celle-ci doit prendre deux vecteurs binaires et associer une valeur dans l'intervalle [0,1] qui quantifie la similitude entre $d_i$ et $d_j$. À cet effet, nous proposons d'utiliser une nouvelle mesure dérivée du concept d'énergie textuelle (Fernández et al., 2007a).

### 3.3 L'énergie textuelle

Soit la matrice document-EL (2), où les valeurs $x_{ji}$ représentent la présence ou l'absence du terme $i$ dans le document $j$. Dans une configuration de réseau d'Hopfield (1982), les valeurs $x_{ji}$ dans la matrice équivalent aux unités du réseau (Fernández et al., 2007, 2007a).

$$X = \begin{vmatrix} x_{11} & x_{12} & \ldots & x_{1p} \\ x_{21} & \ldots & \ldots & \ldots \\ \ldots & \ldots & \ldots & \ldots \\ x_{n1} & \ldots & \ldots & x_{np} \end{vmatrix} \tag{2}$$

L'énergie textuelle d'interaction entre les documents (Fernández et al., 2007) est calculée selon l'équation (3) :

$$E_{textuelle} = -\frac{1}{2}(X \times X^T)^2 \tag{3}$$

où les entrées ($e_{ij}$) de la matrice $E_{textuelle}$ sont toutes négatives ou nulles. Etant donné que l'objectif est de comparer la magnitude de la distance entre vecteurs binaires, sans perte de généralité, on peut considérer les valeurs absolues des entrées de la matrice :

Regroupement sémantique de définitions en espagnol

$$E = \left|-E_{textuelle}\right| \quad (4)$$

Par sa symétrie, $E$ peut être représenté comme un vecteur de distance énergétique (Molina, 2009). Le vecteur (5) contient la distance énergétique entre chaque paire de vecteurs binaires.

$$D_{ener} = [e_{12}, e_{13}, e_{14},..., e_{1n}, e_{23}, e_{24},..., e_{2n},..., e_{n-1n}] \quad (5)$$

Pour restreindre les valeurs des entrées dans l'intervalle [0,1], nous normalisons les entrées $D_{ener}$ en divisant par le maximum.

Nous avons eu l'idée de combiner cette mesure avec le regroupement hiérarchique afin de créer des regroupements des définitions.

## 4  Le corpus de termes polysémiques en espagnol

Le CD est une structure discursive récemment étudiée et les corpus disponibles pour cette tâche sont rares voire inexistants. Nous avons eu donc besoin de créer notre propre corpus. Ses caractéristiques nécessaires pour les expériences sont très précises : d'un côté, il faut prendre en compte la quantité d'acceptions qu'un terme peut avoir selon son contexte (polysémie). D'un autre côté, il faut extraire d'Internet un nombre suffisant des données. Pour ces raisons, on a sélectionné minutieusement les termes à inclure dans le corpus des termes polysémiques en espagnol (CTPE). Selon les indications d'Alameda et Cuetos (1995), on a choisi d'abord une liste $L$ de dix termes ayant les caractéristiques nécessaires : L={*aguja*, *barra*, *cabeza*, *casco*, *célula*, *golpe*, *punto*, *serie*, *tabla*, *ventana* (*aiguille*, *barre*, *tête*, *casque*, *cellule*, *coup*, *point*, *série*, *table*, *fenêtre*)}. Une fois la listé établit, le problème a été de trouver les CDs sur Internet. Ainsi on a associé la liste $L$ à une liste de patrons définitoires afin de former un patron de recherche (PR). Un exemple de PR qui associe le terme *aguja* et le PD *ser+determinante* (*être + déterminant*) est : *la aguja es un* (*l'aiguille est un*). La figure 1 illustre quelques PRs avec le symbole <T> qui représente un terme générique.

```
la <T> est le              définit une <T>
la <T> est la              ...
la <T> est un              nous définissons une <T>
les <T>s sont des          a défini la <T>
...                        a défini une <T>
nous considérons la <T>    ...
```

FIG. 1 – *Patrons de recherche.*

Nous avons utilisé les PRs avec un Web service (L'API BOSS de Yahoo!) afin d'extraire l'information nécessaire. La précision a été restreinte à cause des limitations du Web service. Il est possible -et très courant- de trouver des fragments textuels avec un PD mais qui ne sont pas de définitions. Par exemple, dans le fragment : *En general, el miedo a la aguja es el más frecuente,* (*En générale, la peur de l'aiguille est la plus fréquente*) on observe que le terme *aguja* est présent ainsi que le patron *ser+determinante*, mais évidemment il ne s'agit pas d'une définition. Un fragment de texte est un candidat à contexte définitoire (CCD) s'il contient un terme et un PD mais il n'est pas une définition (Sierra et Alarcón 2002). La très grande quantité d'informations récupérées (environ 3,000 résultats par terme), s'opposait aux critères d'évaluation de l'étude (lecture et interprétation directe). La table 1 montre la quantité de CCDs extraits sur Internet pour chaque terme et genre de définition. À cause de la grande quantité de bruit obtenu nous avons décidé de réduire le nombre de termes. Les termes finalement retenus sont : *barra*, *célula*, *punto*, *ventana*.

|  | **Analytiques** | **Extensionnelles** | **Fonctionnelles** |  |
|---|---|---|---|---|
| **Barra** (*Barre*) | 1.863 | 307 | 467 | **2.637** |
| **Célula** (*Cellule*) | 5.352 | 649 | 533 | **6.534** |
| **Punto** (*Point*) | 1.702 | 422 | 750 | **2.874** |
| **Ventana** (*Fenêtre*) | 1.534 | 587 | 565 | **2.686** |
|  | **10.451** | **1.965** | **2.315** | **14.731** |

TAB. 1 – *CCDs extraits du Web*.

## 5 Evaluation

### 5.1 Méthodologie d'évaluation

L'algorithme de regroupement a été exécuté pour chaque paire (terme-type de définition). Nous avons réalisé un balayage du seuil par distance de 0,01 jusqu'à 1,00, avec un pas de 0,01. Nous avons affiché uniquement les groupes réunissant au moins deux définitions, en éliminant ceux possédant une seule. Afin que le lecteur puisse observer toutes les définitions originellement présentes lors des expériences, nous avons inclus le groupe absolu dont la valeur du seuil par distance est de 1,00.

L'analyse qualitative a consisté à une lecture et interprétation directe des groupes générés, effectuée par un évaluateur humain. Les résultats peuvent être consultés sur le site web http://saussure.iingen.unam.mx/~amolinav/resultados

Afin de comparer les résultats avec ceux de la distance énergétique, nous avons utilisé une adaptation de la distance de Hamming (1950) comme *baseline*. Nous avons divisé le nombre d'entrées différentes entre la longueur des vecteurs afin d'obtenir des valeurs dans la plage [0,1]. Nous présentons les courbes en utilisant les deux critères de distance : l'énergie textuelle et la distance de Hamming.

Nous avons évalué la qualité des regroupements avec trois mesures : le nombre de groupes, la précision et le rappel. À notre connaissance, il n'existe pas une mesure unifiant les trois critères. Même si la *F*-mesure combine la précision et le rappel (Van Rijsbergen, 1979), nous avons décidé de les isoler, car l'adaptation de la précision définie ci-dessous implique la lecture et une interprétation humaine. Le calcul du rappel étant complètement automatique, leur combinaison pourrait donc fausser les résultats.

### 5.2 Nombre de groupes

Après avoir utilisé la distance énergétique, le nombre de groupes augmente en proportion de la valeur de seuil par distance. Le type de courbe est pratiquement le même pour tous les types de définition. Nous avons déduit que le nombre de groupes générés est indépendant du type de définition. La figure 2 illustre le comportement du nombre de groupes en fonction de la valeur de seuil par distance pour les définitions extensionnelles.

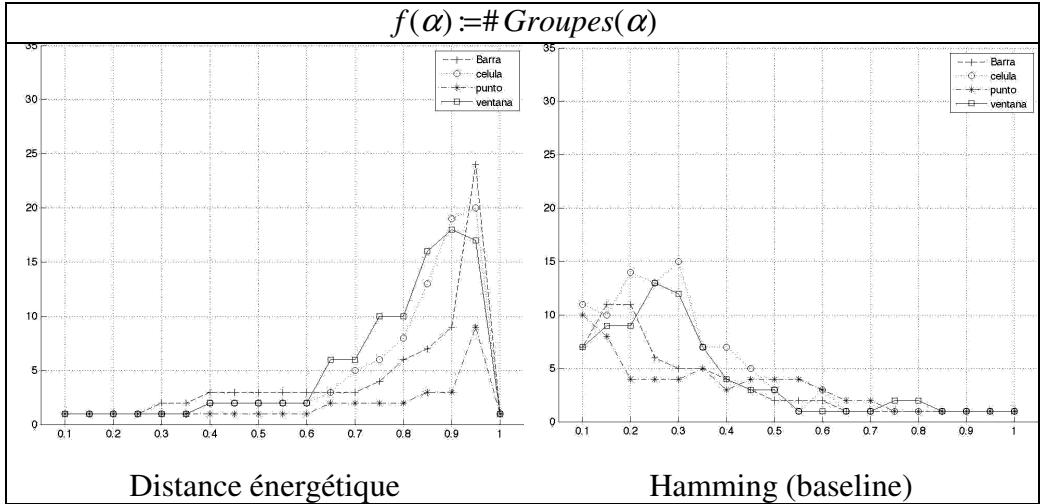

FIG. 2 – *Nombre de groupes en fonction de la valeur de seuil par distance pour les définitions extensionnelles.*

## 5.3 Evaluation du rappel et de la précision

Dans la recherche d'information, le rappel est la proportion de documents récupérés dans une recherche. Nous avons adapté cette définition de rappel comme la proportion de définitions intégrées à un groupe par rapport au total des définitions. C'est-à-dire, combien de définitions nous réussissons à intégrer dans un regroupement. Le rappel prendra la valeur 0 si aucun groupe, avec au moins deux définitions, n'a été identifié ; et 1 si toutes les définitions ont été intégrées dans un groupe. La formule utilisée pour calculer le rappel est la suivante :

$$r = \frac{|\{Totalité\_de\_définitions\} \cap \{Définitions\_inclues\_dans\_un\_groupe\}|}{|\{Totalité\_de\_définitions\}|} \quad (6)$$

La précision est définie comme la fraction de documents pertinents parmi les documents récupérés dans une recherche. Nous avons aussi adaptée cette définition pour quantifier la proportion d'intrus dans le regroupement généré. Elle indique combien d'erreurs nous avons commises après avoir généré les groupes. La précision obtient valeur de 0 si aucun groupe définit auparavant est formé, et de 1 si aucun groupe contient d'intrus. La formule utilisée pour la précision est :

$$p = \frac{|\{Définitions\_inclues\_dans\_un\_groupe\} - \{Intrus\}|}{|\{Définitions\_inclues\_dans\_un\_groupe\}|} \quad (7)$$

Le comportement de la précision et du rappel est indépendant du type de définition. Pour des raisons de clarté nous montrons seulement les résultats des définitions extensionnelles. La figure 3 illustre le comportement de la précision en fonction de la valeur du seuil par distance et la 4 illustre celui du rappel.

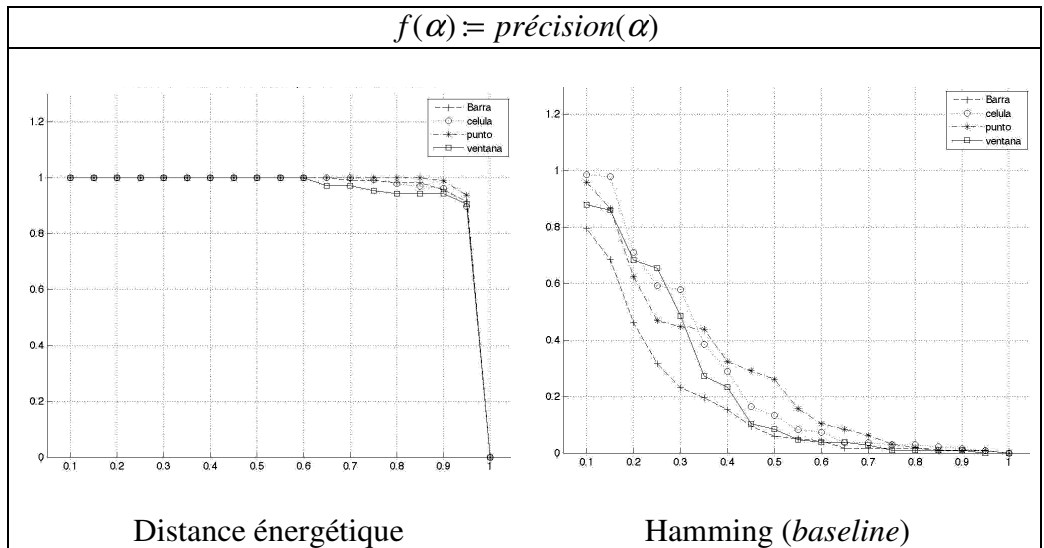

FIG. 3 – *Précision en fonction de la valeur du seuil par distance pour les définitions extensionnelles.*

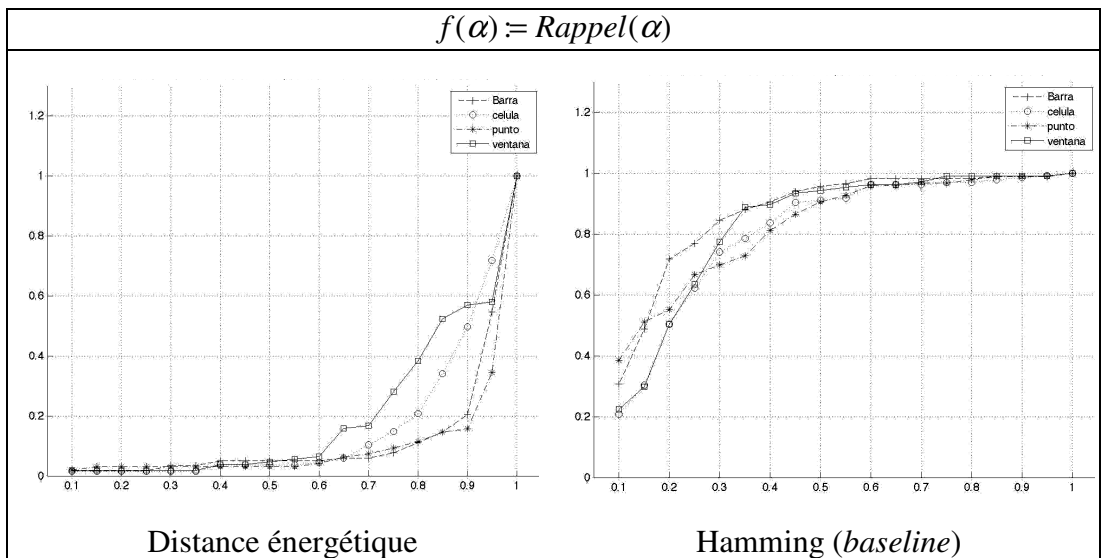

FIG. 4 – *Rappel en fonction de la valeur du seuil par distance pour les définitions extensionnelles.*

## 5.4 Une analyse qualitative

En général, on valide des regroupements qui reflètent les différentes acceptions d'un terme, mais aussi, on distingue des subtilités existantes dans les définitions même à l'intérieur d'une acception identique. Nous pouvons affirmer que l'algorithme regroupe des définitions minutieusement, car il réussit à regrouper des définitions structurellement distinctes mais avec une signification équivalente. L'analyse qualitative semble ainsi très encourageante. Considérez comme exemple de ce phénomène les deux définitions suivantes de *célula* (cellule):

```
La célula se compone de un núcleo envuelto en protoplasma, alrededor
del cual hay una membrana que separa la célula de su medio ambiente[4].
```

```
La célula consta de una membrana celular que envuelve una masa visco-
sa y granulosa llamada protoplasma, en la cual se encuentran todos
los organelos celulares, incluido el núcleo[5].
```

Dans la première, nous pouvons déduire que la cellule est composée d'un noyau qui est enveloppé d'un protoplasme à son tour enveloppé d'une membrane. La deuxième définition est structurellement inversée, mais sémantiquement équivalente : la cellule est composée d'une membrane qui enveloppe le protoplasme où l'on trouve le noyau.

Un autre résultat important de l'analyse est le suivant : à mesure que la valeur de seuil tend vers 1, les groupes deviennent plus spécifiques et parfois inconvenablement explicites dans la signification des mots qui les conforment. Nous avons aussi observé que lorsque plus de définitions sont incluses dans le regroupement, un bruit fort s'introduit au sein de groupes qui ont été crées. Nous parlons des intrus qui apparaissent dans les groupes et qui sont incongrus avec la majorité des définitions du groupe.

## 5.5 Analyse quantitative

Nous avons constaté que le comportement de l'algorithme est indépendant du type de définition. Cela montre que l'algorithme peut être utilisé pour n'importe quel type de texte et pas seulement pour des définitions.

Le nombre de groupes, la précision et le rappel sont proportionnels au seuil. Il faut mentionner que, mis à part le type de définition, le comportement de l'algorithme peut être divisé en zones par rapport à la valeur de seuil par distance :

1.  La zone 1 où $0 \leq \alpha \leq 0.7$ : dans cette zone on obtient une très bonne précision (supérieure au 90%) et un rappel bas (inférieur à 40%). Il faut utiliser une valeur α dans cet intervalle pour obtenir des acceptions courantes, un nombre réduit de groupes (environ 5) avec peu de définitions et sans la présence d'intrus dans les groupes générés.

---

[4] La cellule est composée d'un noyau enveloppé dans le protoplasme, autour duquel il y a une membrane qui sépare la cellule de son milieu.
[5] La cellule est composée d'une membrane cellulaire qui enveloppe une masse visqueuse et granuleuse appelée le protoplasme, dans lequel on trouve toutes les organelles cellulaires, y compris le noyau.

2. La zone 2 où $0.75 \leq \alpha \leq 0.85$ se caractérise par un haut degré de précision (autour de 80%) et un rappel moyen (autour de 50%). Cet intervalle maintient un bon équilibre entre la précision et le nombre de groupes générés (environ 10) ainsi qu'un rappel acceptable.
3. La zone 3 où $0.85 \leq \alpha \leq 0.99$, obtient une précision moyenne (environ 50%) et un rappel élevé (environ 80%). Le nombre de groupes générés dans cette zone est trop élevé (supérieur à 20) -il y a plus de groupes que d'acceptions-, par contre chaque groupe est très précis en termes de la cohérence de sa signification.

## 6 Conclusions et travail futur

Même si ce travail a été testé sur un corpus de définitions en espagnol, la méthode présentée est indépendante des considérations linguistiques. Une étude d'adaptation à la langue française est en cours au Laboratoire Informatique d'Avignon. En ce qui concerne la complexité, il faut considérer l'espace mémoire (en raison de la génération de la matrice document-EL qui est toujours d'ordre $O(P^2)$, où $P$ est le nombre de documents. Ainsi, la méthode est bien adaptée pour regrouper des textes courts (quelques dizaines de phrases). Par exemple des extraits de résultats des moteurs de recherche (*snippets*), les manchettes, les résumés automatiques, etc.

Une amélioration possible de notre méthode de regroupement de définitions consiste à calculer dynamiquement la valeur de seuil grâce à laquelle on obtiendrait le meilleur regroupement en prenant en compte les caractéristiques désirées (nombre de groupes précis, haute précision, rappel élevé, etc.). On pourrait aussi fixer la valeur du seuil α afin de comparer des distances binaires différentes. Ces études son actuellement en cours.

## Remerciements

## Références

Regroupement sémantique de définitions en espagnol